\documentclass[pra,showpacs,twocolumn,nofootinbib,aps]{revtex4}

\usepackage{amsmath}
\usepackage{amssymb}
\usepackage{latexsym}
\usepackage{amsfonts}
\usepackage{epsfig}
\usepackage{graphicx}
\usepackage{subfigure}
\usepackage{hyperref}

\usepackage{epsfig}
\usepackage[utf8]{inputenc}

\newcommand{\ket}[1]{\left|#1\right>}

\newcommand{\f}[1]{\mbox{\boldmath$#1$}}
\newcommand{\fk}[1]{\mbox{\boldmath$\scriptstyle#1$}}

\newcommand{\bea}{\begin{eqnarray}}
\newcommand{\ea}{\end{eqnarray}}
\newcommand{\eea}{\end{eqnarray}}
\newcommand{\ord}{\,{\cal O}}

\newcommand{\ann}[1]{\hat{a}_{\fk{#1}}}

\newcommand{\summ}{\sum_{\mu\nu} T_{\mu \nu}}
\newcommand{\bcre}[1]{\hat{b}_{#1}^\dagger}
\newcommand{\bann}[1]{\hat{b}_{#1}}
\newcommand{\ccre}[1]{\hat{c}_{#1}^\dagger}
\newcommand{\cann}[1]{\hat{c}_{#1}}
\newcommand{\kin}{\fk{\kappa}_{\rm in}}
\newcommand{\Kin}{\f{\kappa}_{\rm in}}
\newcommand{\kout}{\fk{\kappa}_{\rm out}}
\newcommand{\Kout}{\f{\kappa}_{\rm out}}

\begin{document}

\title{Dicke superradiance as a nondestructive probe for quantum quenches 
in optical lattices}

\author{Nicolai ten Brinke and Ralf Sch\"utzhold}

\email{ralf.schuetzhold@uni-due.de}

\affiliation{Fakult\"at f\"ur Physik, Universit\"at Duisburg-Essen, 
Lotharstrasse 1, D-47057 Duisburg, Germany}

\date{\today}

\begin{abstract}
We study Dicke superradiance as collective and coherent absorption and 
(time-delayed)
emission of photons from an ensemble of ultracold atoms in an optical 
lattice.
Since this process depends on the coherence properties of the atoms 
(e.g., superfluidity), it can be used as a probe for their quantum state.
In analogy to pump-probe spectroscopy in solid-state physics, 
this detection method facilitates the investigation of nonequilibrium 
phenomena and is less invasive than time-of-flight experiments or 
direct (projective) measurements of the atom number (or parity) per lattice 
site, which both destroy properties of the quantum state such as phase 
coherence. 
\end{abstract}

\pacs{42.50.Gy, 05.30.Jp, 03.75.Lm.}

\maketitle

\section{Introduction}

Ultracold atoms in optical lattices are very nice tools for investigating 
quantum many-body physics since they can be well isolated from the 
environment and cooled down to very low temperatures 
\cite{Raizen:1997yq,Bloch:2004eg,Bloch:2005zl}.
Furthermore, it is possible to control these systems and to measure their 
properties to a degree which cannot be reached in many other scenarios.
For example, the quantum phase transition \cite{Sachdev:2001kx} 
between the highly correlated Mott insulator state and the superfluid phase 
in the Bose-Hubbard model \cite{Fisher:1989fv,Jaksch:1998uq,Jaksch:2005kq} 
has been observed \cite{Greiner:2002fk,Stoferle:2004qv,Spielman:2007kx}. 
This observation was accomplished by time-of-flight experiments where the 
optical lattice trapping the atoms is switched off and their positions are 
measured after a waiting time.  
As another option for detecting the state of the atoms, the direct 
{\em in situ} measurement of the number of atoms per lattice site
(or more precisely, the parity) has been achieved recently (see, e.g.,  
\cite{Sherson:2010uq,Endres:2013fk}).
However, both methods are quite invasive since they destroy properties of 
the quantum state such as phase coherence%
\footnote{In addition, one has to be careful in interpreting the momentum
distribution of a time-of-flight measurement, as it was, e.g., shown 
that sharp peaks are not a reliable witness of superfluidity
\cite{Kato:2008vn}.}. 

Methods for less destructive probing of the quantum state of atoms in optical 
lattices were proposed recently, e.g., the interaction with light in an 
optical cavity 
\cite{Landig:2015aa,Rajaram:2013zl,Silver:2010rz,Bhaseen:2009fv,Zoubi:2009ty,Chen:2007aa,Mekhov:2007}
or matter-wave scattering with (slow) atoms 
\cite{Sanders:2010aa,Mayer:2014aa}.
In this paper, we study an alternative, nondestructive detection method%
\footnote{
In contrast to instantaneous off-resonant Bragg-type scattering of 
cavity or resonator modes considered previously \cite{Mekhov:2007,Chen:2007aa},
or vacuum-stimulated scattering of light to directly measure the 
dynamic structure factor \cite{Landig:2015aa},
we study resonant Dicke superradiance in free space with a time delay 
between absorption and emission.
As a result, our method is sensitive to the correlator of creation 
$\hat{b}_\mu^\dagger(t)$ and annihilation $\hat{b}_\nu(t')$ 
operators~(\ref{eq:eightpoint}) including their phase coherence 
at different times $t$ and $t'$ (and lattices sites $\mu$ and $\nu$) 
instead of the correlator containing on-site number operators 
$\hat{n}_\nu(t)$ at the same time $t$ only, as in \cite{Mekhov:2007}. 
Employing the analogy to solid-state physics, these previous approaches are 
similar to Bragg scattering (Debye-Waller factor, etc.), whereas our method 
corresponds to pump-probe spectroscopy with a time delay -- which provides 
important complementary information, e.g., for nonequilibrium phenomena. 
}
based on Dicke superradiance, i.e., the (free-space) collective and coherent 
absorption and emission of photons from an ensemble of ultracold atoms 
(see, e.g., \cite{Dicke:1954kx,Rehler:1971fk,Lipkin:2002fk,Wiegner:2011aa}).
We investigate how the lattice dynamics (e.g., hopping) occurring between 
the absorption and subsequent superradiant emission of single photons 
\cite{Scully:2006fk,Scully:2007fk,Oliveira:2014aa} changes the emission 
characteristics.

\begin{figure}[h]
\begin{center}
\includegraphics[width=0.8\columnwidth]{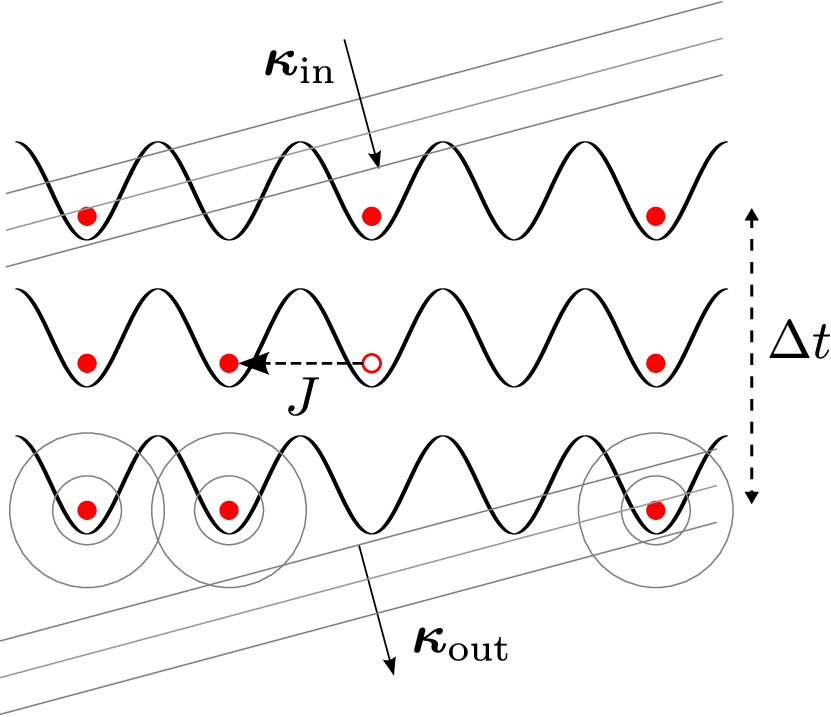}
\caption{(Color online)
 Envisaged probing sequence.
First, an infrared photon $\Kin$ is absorbed collectively
by the ground-state atoms at their respective lattice sites.
Second, the atoms tunnel and interact according to (\ref{eq:BHHam}) 
during a waiting period $\Delta t$, thereby possibly compromising
the spatial phase coherence of the Dicke state.
In the third step, the photon is (collectively) emitted again with the
wave vector $\Kout$.
}
\label{fig:probe_dicke}
\end{center}
\end{figure}

To probe the quantum state of the optical lattice, we envisage the
following sequence (see Fig.~\ref{fig:probe_dicke}):
First, an infrared photon is absorbed by one of the atoms,
but we do not know which one (creating a ``timed'' Dicke state 
\cite{Scully:2006fk,Scully:2007fk,Oliveira:2014aa}).
Then, during a waiting period $\Delta t$, the atoms have time to 
tunnel and to interact.
Afterwards, the excited atoms decay back to their ground state
by collectively emitting an infrared photon -- 
depending on the coherence properties of the atoms. 
Options for experimental realizations will be discussed below. 
%
\section{Basic Formalism}

%

Under appropriate conditions, bosonic ultracold atoms in optical lattices are 
approximately described by the Bose-Hubbard Hamiltonian \cite{Jaksch:2005kq},
\bea
\label{eq:BHHam}
\hat{H}_{\rm BH}
=  
-\frac{J}{Z}\summ 
\bcre{\mu}\bann{\nu}+\frac{U}{2}\sum_\mu\hat{n}_\mu^{(b)}(\hat{n}_\mu^{(b)}-1)
\,,
\ea
with the hopping rate $J$, the interaction strength $U$, 
the adjacency matrix $T_{\mu\nu}$, and the coordination number $Z$.
Here, we assume a quadratic lattice with $Z=4$.
Furthermore, $\bcre{\mu}$ and $\bann{\nu}$ denote the creation and 
annihilation operators of the atoms (in their ground state) at lattice sites 
$\mu$ and $\nu$, respectively, and $\hat{n}_\mu^{(b)}=\bcre{\mu}\bann{\mu}$
is the number operator. 
Assuming unit filling $\langle\hat{n}_\mu^{(b)}\rangle=1$, this model displays 
a quantum phase transition \cite{Sachdev:2001kx} 
between the superfluid phase where $J$
dominates and the Mott insulator state where $U$ dominates. 
In the extremal limits $J\gg U$ and $U\gg J$, the ground states simply read 
\bea
\label{Mott}
\ket{\Psi}_{\rm Mott}^{J=0}=\prod\limits_\mu\bcre{\mu}\ket{0}=
\bigotimes\limits_\mu\ket{1}_\mu
\,,
\ea
for the Mott insulator state and 
\bea
\label{superfluid}
\ket{\Psi}_{\rm superfluid}^{U=0}
=
\frac{1}{\sqrt{N!}}
\left(\hat b_{\fk{k}=0}^\dagger\right)^N\ket{0}
\propto
\left(\sum\limits_\mu\bcre{\mu}\right)^N\ket{0}
\,,
\ea
for the superfluid phase, where $N$ is the total number of lattice sites
(which equals the number of particles). 

Now, we consider the interaction of these atoms with infrared photons.
Assuming that the wavelength of these infrared photons is much larger than 
the lattice spacing of the optical lattice, we have a large number of atoms 
within one (infrared) photon wavelength.
In addition, the atomic recoil due to the absorption or emission of an 
infrared photon is negligible, which is another basic requirement for 
Dicke superradiance.  
The interaction between atoms and photons is then described by 
\bea
\label{eq:hInt}
\hat{H}_{\rm int} = 
\int d^3k\, g_{\fk{k}}(t) \ann{k} 
\sum_\mu \ccre{\mu}\bann{\mu}\exp\left(i\f{k}\cdot\f{r}_\mu\right)
+ {\rm H.c.},
\ea
where $\ann{k}$ is the annihilation operator of a photon with wave number 
$\f{k}$ and $\f{r}_\mu$ is the position of the atom at the lattice site 
$\mu$.
The excited atoms at lattice sites $\mu$ and $\nu$ are described by the 
creation and annihilation operators $\ccre{\mu}$ and $\cann{\nu}$ and thus 
one has to extend the Bose-Hubbard Hamiltonian~(\ref{eq:BHHam}) accordingly. 
%

\section{Emission Probability}

In lowest-order perturbation theory, the probability (density) to first absorb 
a photon with wave number $\Kin$ and later (after the waiting time $\Delta t$)
emit a photon with wave-number $\Kout$ reads  
\bea
\label{probability}
P
&=&
\int dt_1\,dt_2\,dt_3\,dt_4\,
g_{\rm out}(t_4) g_{\rm out}^*(t_2) g_{\rm in}^*(t_3) g_{\rm in}(t_1)
\nonumber\\
&&\times 
\exp\left\{i(\omega_{\rm in}t_3-\omega_{\rm out}t_4)
-i(\omega_{\rm in}t_1-\omega_{\rm out}t_2)\right\}
\nonumber\\
&&\times
\mathcal{D}\left(t_1,t_2,t_3,t_4\right) 
\,,
\ea
with the operatorial part containing the lattice dynamics 
\bea
\label{eq:eightpoint}
\mathcal{D}\left(t_1,t_2,t_3,t_4\right) 
=
\sum_{\mu\nu\rho\eta} 
\exp\left\{i\left(\Kout\cdot\f{r}_\rho-\Kin\cdot\f{r}_\eta\right)\right\}
\nonumber\\
\times
\exp\left\{-i\left(\Kout\cdot\f{r}_\mu-\Kin\cdot\f{r}_\nu\right)\right\}
\nonumber\\
\times
\langle 
\hat{b}_\eta^\dagger(t_3)\hat{c}_\eta(t_3) 
\hat{c}_\rho^\dagger(t_4)\hat{b}_\rho(t_4) 
\hat{b}_\mu^\dagger(t_2)\hat{c}_\mu(t_2)
\hat{c}_\nu^\dagger(t_1)\hat{b}_\nu(t_1)
\rangle 
.
\ea
The expectation value in the last line should be taken in the initial state,
which could be a pure state, such as the superfluid state~(\ref{superfluid})
or the Mott state~(\ref{Mott}), or a mixed state such as a thermal density 
matrix.  

As a result, this probability (density) depends on the above four-times 
eight-point function, which contains information about the underlying state.  
Unfortunately, since the Bose-Hubbard model is 
not integrable, we do not have an explicit solution for this eight-point 
function apart from some limiting cases. 
Let us first study the special case that the initial state 
$\hat\varrho^{\rm in}$ 
contains no correlations and zero ($\hat{c}_\mu^\dagger$) excitations, 
i.e., it can be represented by a product of single-site states 
$\hat\varrho^{\rm in}=\otimes_\mu\hat\varrho^{\rm in}_\mu$ with  
$\langle\hat{c}_\mu^\dagger\hat{c}_\nu\rangle=0$.
Furthermore, we assume that the correlations which arise through
the time evolution $\Delta t$ remain negligible. Then, the operatorial
part reads to leading order in $N$,
\begin{multline}
\label{eq:NoCorr}
\mathcal{D}_{\kout}^{\kin}\left(t,t'\right) 
=
\Big|\sum_{\mu} 
\exp\left\{-i\left(\Kout-\Kin\right)\cdot\f{r}_\mu\right\}\times
\\
\times
\langle
\hat{b}_\mu^\dagger(t') \hat{c}_\mu(t') \hat{c}_\mu^\dagger(t) \hat{b}_\mu(t)
\rangle
\Big|^2
\,,
\end{multline}
where $t=t_{1/3}$ denotes the time when the photon is absorbed and 
$t'=t_{2/4}$ the emission time. 
The result is quite intuitive, as the above expectation value is just the 
probability amplitude that the excited atom stays at the lattice site 
$\mu$ during the waiting time $\Delta t = t'-t$.
In the limit of $J=0$, the atoms are pinned to their lattice sites 
and we get the usual Dicke superradiance.
In this case, the expectation value gives the number $n_\mu$ of atoms 
at site $\mu$, and the sum can be interpreted as a discrete Fourier 
transform of the $n_\mu$-distribution of the atoms in the lattice.
Considering, e.g., the Mott state (\ref{Mott}), where all the atoms
are equally distributed, the Fourier transform yields a sharp peak 
$\delta(\Kin,\Kout)$, which corresponds to the well-known
directed spontaneous (superradiant) emission for fixed atoms 
\cite{Scully:2006fk,Scully:2007fk,Oliveira:2014aa}.
However, we would like to stress again that Eq.~(\ref{eq:NoCorr}) is only 
valid when the correlations between lattice sites are negligible. 
Turning this argument around, a deviation from Eq.~(\ref{eq:NoCorr})
is an indicator for correlations. 

In the other limiting case $U=0$, we may also simplify 
Eq.~(\ref{eq:eightpoint}). 
Assuming that there are no excited atoms initially 
$\langle\hat{c}_\mu^\dagger\hat{c}_\nu\rangle=0$, the eight-point function 
in (\ref{eq:eightpoint}) can be reduced to a four-point function in terms 
of the operators $\bcre{\mu}$ and $\bann{\nu}$.
After a Fourier transform, this four-point function 
$\langle\bcre{\fk{q}-\kin}\bann{\fk{q}-\kout}\bcre{\fk{p}-\kout}
\bann{\fk{p}-\kin}\rangle$ depends on two wave numbers $\f{p}$ and $\f{q}$
(assuming translational invariance). 
If we have a Gaussian state (for $U=0$), such as a thermal state or 
(to a very good approximation) the superfluid state~(\ref{superfluid}), 
it can be expanded into a sum of products of two-point functions via the 
Wick theorem.
Finally, if the initial state is diagonal in the $\f{k}$ basis -- 
which is also the case for the superfluid state~(\ref{superfluid}) 
and thermal states -- these two-point functions just give the spectrum 
$N_{\fk{k}}$, i.e., the number of particles per mode $\f{k}$.
For example, the expectation value 
$\langle\bcre{\fk{q}-\kin}\bann{\fk{q}-\kout}\rangle$ becomes 
$N_{\fk{q}-\kin}\delta(\Kin,\Kout)$.  

\section{Super-Radiance}\label{Sec-Super-Radiance}

As an example for the general considerations above, let us consider the 
superfluid state~(\ref{superfluid}) with $U=0$ as the initial state.
In this case, the probability (density) in Eq.~(\ref{probability}) is 
independent of the waiting time $\Delta t$ and yields  
\bea
\label{eq:probsf}
P=N^2\delta(\Kin,\Kout)P_{\rm single}
\,,
\ea
to leading order, where $P_{\rm single}$ is the corresponding expression for 
a single atom. 
As a result, we obtain the same Dicke superradiance as in the case of 
immovable atoms. 
Note that one factor of $N$ originates from the simple fact that $N$ atoms 
absorb the incident photon more likely than one atom -- whereas the other 
factor of $N$ corresponds to the coherent enhancement of the collective 
decay probability (i.e., Dicke superradiance). 

As the next example, let us consider a state where $N_1$ atoms are in the 
superfluid state (with $\f{k}=0$) while the other $N_2$ atoms are equally 
distributed over all $\f{k}$ modes. 
This can be considered as a simple toy model for a thermal state with partial 
condensation, for example. 
In this situation, the probability (density) in Eq.~(\ref{probability}) does 
depend on the  waiting time $\Delta t$ and behaves as 
\bea
\label{eq:probability}
P=\left|N_1 e^{i\varphi(\Delta t)}+N_2{\mathcal J}(\Delta t)\right|^2
\delta(\Kin,\Kout)P_{\rm single}
\,.
\ea
The phase $\varphi(\Delta t)$ can lead to interference effects between the 
two terms and  is given by $\varphi(\Delta t)=J(T_{\fk{\kappa}}-1)\Delta t$, 
where we have abbreviated $\f{\kappa}=\Kin=\Kout$ and $T_{\fk{k}}$ 
denotes the Fourier transform of the adjacency matrix $T_{\mu\nu}$.  
For a quadratic lattice with lattice spacing $\ell$, it reads 
$T_{\fk{k}}=[\cos(k_x\ell)+\cos(k_y\ell)]/2$.
The remaining function ${\mathcal J}(\Delta t)$ describes the reduction of 
superradiance due to the hopping of the excited atoms during the waiting time,
\bea
\label{reduction}
\mathcal{J}(\Delta t)
=
\frac1N\sum_{\fk{k}}
\exp\left\{iJ(T_{\fk{k}}-T_{\fk{k}-\fk{\kappa}})\Delta t\right\}
\leq1
\,.
\ea
For small wave numbers $|\f{\kappa}|\ell\ll1$ and large enough waiting 
times $\Delta t$ such that $J\Delta t|\f{\kappa}|\ell=\ord(1)$, it can be 
approximated by Bessel functions $J_0$,
\bea
\label{eq:bessels}
\mathcal{J}(\Delta t)
\approx
J_0 \left( \frac{J\Delta t}{2}\,\kappa_x\ell \right) 
J_0 \left( \frac{J\Delta t}{2}\,\kappa_y\ell \right) 
\,. 
\ea
As a result, the peak in forward direction decays with time $\Delta t$ 
unless the photon was incident in orthogonal direction $\kappa_x=\kappa_y=0$ 
or all atoms are in the superfluid state ($\f{k}=0$), i.e., $N_2=0$. 
This can be explained by the fact that the (excited) atoms tunnel during 
the waiting time $\Delta t$ and thus the initial and final phases 
$\exp\left(i\f{k}\cdot\f{r}_\mu\right)$ do not match anymore. 
The explicit dependence of the Bessel functions on the wave-vector 
$\f{\kappa}$ is a clear deviation from Eq.~(\ref{eq:NoCorr}) and 
demonstrates the significance of correlations between lattice sites, 
which are induced by the hopping $J$.
In summary, a fully condensed state ($N_2=0$) can be distinguished from a 
partially exited (e.g., thermal) gas of atoms ($N_2\neq0$) via Dicke 
superradiance. 

\section{Phase Transition}

Now, after having discussed the two cases $J=0$ and $U=0$ separately, let us 
consider a phase transition between the two regimes. 
After the initial Mott state~(\ref{Mott}) has absorbed the incident photon 
with wave-number $\Kin$, we have the following excited state 
\bea
\label{Mott-excited}
\ket{\Psi}_{\rm excited}^{\rm Mott}
=
\frac{1}{\sqrt{N}}
\sum_\mu \exp\left(i\Kin\cdot\f{r}_\mu\right) \ccre{\mu}
\prod\limits_{\nu\neq\mu}\bcre{\nu}\ket{0}
\,.
\ea
Actually, for $U\gg J$, this is an approximate eigenstate of the 
Bose-Hubbard Hamiltonian~(\ref{eq:BHHam}) in the subspace where 
one atom is excited and the $N-1$ others are not.  
Now, assuming that $N$ is large but finite, we could envisage an adiabatic 
transition from the initial Mott regime $U\gg J$ to the superfluid phase 
where $J\gg U$.
Due to the adiabatic theorem, an initial eigenstate such as the 
state~(\ref{Mott-excited}) stays an eigenstate during that evolution and 
thus we end up with the state (for $N\gg1$),
\bea
\label{sf-excited}
\ket{\Psi}_{\rm excited}^{\rm superfluid}
=
\frac{1}{\sqrt{(N-1)!}}\,
\left(\hat b_{\fk{k}=0}^\dagger\right)^{N-1}\,\hat c_{\kin}^\dagger
\ket{0}
\,,
\ea
where the excited atom possesses the initial wave number $\Kin$ of the 
absorbed photon and the $N-1$ other atoms are in the superfluid 
state~(\ref{superfluid}).
Calculating the emission probability from this state, we find that it shows 
precisely the same characteristic features of Dicke superradiance and thus 
photons are emitted predominantly in the $\Kin$ direction (as one would expect).

As the opposite limit to an adiabatic passage from the Mott state to the 
superfluid phase, let us study the sudden switching procedure
(quantum quench). 
Again starting in the state~(\ref{Mott-excited}), we now envisage an abrupt
change from $J=0$ (and $U>0$) to $U=0$ (and $J>0$).
After this sudden switch, the state~(\ref{Mott-excited}) is no longer an 
eigenstate of the Bose-Hubbard Hamiltonian~(\ref{eq:BHHam}) but a mixture 
of excited states.
Calculating the emission probability from this state, we find that it 
coincides with Eqs.~(\ref{eq:probability}) and (\ref{reduction}) 
for $N_2=N$ and $N_1=0$.
Ergo, the initial Mott state -- after the sudden switch -- behaves as a state 
where all momenta are equally populated. 
This is a quite intuitive result, but one should keep in mind that the 
state~(\ref{Mott-excited}) is not a Gaussian state such that some care is 
required by applying the results from the previous section.
Nevertheless, one can distinguish an adiabatic from a sudden transition via 
Dicke superradiance as the emission characteristics are different. 

\section{Experimental realization}
%
Let us now discuss possible experimental realizations.
We consider an optical lattice formed by a green 
(e.g., argon-ion \cite{Bridges:1964ly}) 
laser with $\lambda_{\rm lat}=514\;{\rm nm}$ with a lattice constant  
$\ell=\lambda/2=257\;{\rm nm}$
\cite{Stamper-Kurn:1998uq,Inouye:1998fk,Jaksch:1998uq,Jaksch:2005kq}.
If we assume that the incoming photon has a wavelength of
$\lambda_{\rm photon}=2\pi/|\Kin|=10.6\;\mu{\rm m}$ \cite{Patel:1964fk},
the recoil energy of the incoming infrared photon is a factor 
$E_R^{\rm lat}/E_R^{\rm photon} = 4\cdot10^2$ smaller than the 
recoil energy of an optical lattice photon, thus the atomic recoil due to the 
absorption or emission of the infrared photon is negligible.
Furthermore, the ratio $\ell^2/\lambda_{\rm photon}^2$ is small enough to ensure 
that collective coherent emission (i.e., Dicke superradiance) is faster than
spontaneous incoherent emission of single atoms (see, e.g., 
\cite{Brinke:2013fk}).
For this reason, the absorbed and emitted photon needs to be in the infrared 
region.
In the following, we specify three options for infrared emission.

{\bf I.} The most straightforward way to implement the probing sequence 
displayed in Fig.~\ref{fig:probe_dicke} would be a simple two-level system 
with an infrared transition and a life-time which is sufficiently long 
compared to the time scales of the lattice dynamics 
(e.g., the tunneling time of typically 
$\tau_{\rm tunnel} = \hbar/J = 5\cdot10^{-5}\;{\rm s}$).
This seems hard to achieve with the usual atoms (e.g., Rb, Na) 
used in optical lattices, but may be feasible using molecules.

\begin{figure}[h]
\begin{center}
\subfigure[]{\includegraphics[width=0.4\columnwidth]{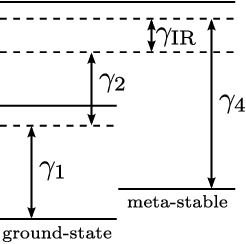}}
\hspace{.5cm}
\subfigure[]{\includegraphics[width=0.4\columnwidth]{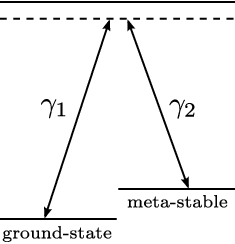}}
\caption{Sketches (not to scale) of the proposed level schemes for the 
experimental implementation.
}
\label{fig:infrared_and_classical}
\end{center}
\end{figure}

{\bf IIa.} This motivates replacing the single-photon transition envisaged 
above by multiphoton transitions. 
For example, one could imagine a detuned four-photon transition as depicted 
in Fig.~\ref{fig:infrared_and_classical}(a), where three participating 
photons $\gamma_1$, $\gamma_2$, and $\gamma_4$ are provided by external 
lasers while the the fourth missing photon $\gamma_{\rm IR}$ is the infrared 
photon under consideration. 
This scheme further facilitates controlling the involved time scales since 
the three external lasers $\gamma_1$, $\gamma_2$, and $\gamma_4$ 
can be switched on 
(during absorption and emission of the infrared photon $\gamma_{\rm IR}$)
and off 
(during the waiting time $\Delta t$) by hand. 

{\bf IIb.} Alternatively, the Dicke state can also be generated by a 
two-photon process as sketched in Fig.~\ref{fig:infrared_and_classical}(b), 
where one photon $\gamma_1$ is provided by an incident laser.
Detecting the momentum $\f{k}_{\rm Stokes}$ of the emitted (or scattered) 
Stokes photon then yields $\Kin$ (see, e.g., 
\cite{Scully:2006fk,Scully:2007fk}).

{\bf III.} Instead of a pure Dicke state corresponding to the absorption and 
emission of single photons, i.e., a well-defined number of excitations, we 
could also use coherent states as generated by classical laser fields. 
Let us consider two counter-propagating lasers ($\gamma_1$ and $\gamma_2$) 
which are switched on for a short time such that they excite on average 
a certain number $\bar n$ of atoms via the detuned two-photon (Raman) 
transition in Fig.~\ref{fig:infrared_and_classical}(b).
In terms of the effective angular momentum operators 
$\Sigma_+=\sum_\mu\ccre{\mu}\bann{\mu}\exp(i\Kin\cdot\f{r}_\mu)$  
and 
$\Sigma_-=\Sigma_+^\dagger$, as well as $\Sigma_z=[\Sigma_+,\Sigma_-]/2$, 
this transition corresponds to a simple rotation, where $\Kin$ is given by 
the momentum difference $\f{k}_1-\f{k}_2$ of the lasers. 
For example, if $\bar n$ is smaller than unity, the resulting coherent state 
is well approximated by a coherent superposition of the ground state 
$\ket{\rm ground}$ with $\Sigma_-\ket{\rm ground}=0$ 
and the first excited Dicke state $\Sigma_+\ket{\rm ground}$.  
Then, after a waiting time $\Delta t$, we may switch on the two lasers again 
in order to reverse this rotation. 
If the atoms did not evolve (e.g., tunnel) during that time $\Delta t$,
we would get the ground state $\ket{\rm ground}$ afterwards. 
However, if the atoms tunnel and thereby scramble their spatial phases,
the rotation back to the ground state would not be perfect and we would 
obtain a finite probability for some excited atoms remaining in the final 
state -- which could then be detected. 

For example, if we consider the same initial state as in 
Eq.~(\ref{eq:probability}), where $N_1$ atoms are condensed ($\f{k}=0$)
and the remaining $N_2=N-N_1$ atoms are equally distributed over all other 
$\f{k}$ modes, the average number of exited atoms in the final state reads 
(to leading order)
\bea
\label{eq:probability_classical}
\langle \hat{n}_{\rm meta} \rangle 
=
\bar n\left(
1 - \frac{N_1}{N} 
\cos\left\{\varphi(\Delta t)\right\} - \frac{N_2}{N} {\mathcal J}(\Delta t)
\right)
\,.
\ea
Thus, by measuring $\langle \hat{n}_{\rm meta} \rangle$ as a function of the 
waiting time $\Delta t$, we may infer the number $N_1$ of condensed atoms. 
Note that $\varphi(\Delta t)$ and ${\mathcal J}(\Delta t)$ are exactly the 
same expressions as in Eq.~(\ref{eq:probability}), which shows that the 
two scenarios are very similar.
The most obvious difference is the interference term 
$\cos\{\varphi(\Delta t)\}$ stemming from the fact that we have a coherent  
superposition of states with different energies instead of a pure Dicke state 
(as mentioned above).
However, the different time-dependences -- 
oscillation $\cos\{\varphi(\Delta t)\}$ versus decay ${\mathcal J}(\Delta t)$
-- should allow us to distinguish the two mechanisms. 
Thus, it should also be possible to differentiate between an adiabatic 
passage from the Mott state to the superfluid phase and a sudden transition. 

\section{Conclusions}

We studied Dicke superradiance from an ensemble of ultracold atoms in an 
optical lattice described by the Bose-Hubbard Hamiltonian~(\ref{eq:BHHam})
and found that the character of the emission probability (\ref{probability}) 
can be employed to obtain information about the 
evolution of the quantum state of the atoms.
In the noninteracting case $U=0$, for example, the temporal decay of the 
emission peak in forward direction~(\ref{eq:probability}) and 
(\ref{reduction}) can be used to infer the number $N_1$ of condensed atoms. 
Comparing the adiabatic passage from the Mott state to the superfluid phase 
with a sudden transition, we found that these two cases can also be 
distinguished via the temporal behavior of the emission probability. 
Finally, we discussed several options for an experimental realization.

Note that the above method is complementary to other techniques since it 
yields information about the 
temporal evolution of the coherence properties of the atoms without 
destroying their state.
Analogously to pump-probe spectroscopy in solid-state physics, 
the dependence of (\ref{probability}) on initial and final wave numbers 
$\Kin$ and $\Kout$ as well as on waiting time $\Delta t$ yields 
nonequilibrium spectral information. 
Since Eq.~(\ref{eq:eightpoint}) includes different time coordinates, 
we obtain access to double-time Green functions \cite{Zubarev:1960aa} 
and thus may 
distinguish even- and odd-frequency correlators \cite{Berezinskii:1974aa}, which 
also became a topic of increasing interest recently.
Here, we mainly focused on the Mott--superfluid phase transition in the 
Bose-Hubbard model because it is well studied experimentally, but our 
method can be applied to other cases -- as long as they display distinctive 
signatures in the correlator (\ref{eq:eightpoint}).
For example, the quench from the Mott insulator state to the metallic phase 
in the Fermi-Hubbard model can be studied in an analogous manner. 
One would even expect that superconductivity shows signatures in a correlator 
of the form (\ref{eq:eightpoint}), but our understanding of these matters 
is not complete yet.

\acknowledgments

This work was supported by the DFG (SFB-TR12). 


\bibliographystyle{apsrev}
\bibliography{d12}

\end{document}